# Pitchtron: Towards audiobook generation from ordinary people's voices

*Sunghee Jung, Hoirin Kim*

School of Electrical Engineering, KAIST, Daejeon, Republic of Korea

`{sh.ee,hoirkim}@kaist.ac.kr`

## Abstract

In this paper, we explore prosody transfer for audiobook generation under rather realistic condition where training DB is plain audio mostly from multiple ordinary people and reference audio given during inference is from professional and richer in prosody than training DB. To be specific, we explore transferring Korean dialects and emotive speech even though training set is mostly composed of standard and neutral Korean. We found that under this setting, original global style token method generates undesirable glitches in pitch, energy and pause length. To deal with this issue, we propose two models, hard and soft pitchtron and release the toolkit and corpus that we have developed. Hard pitchtron uses pitch as input to the decoder while soft pitchtron uses pitch as input to the prosody encoder. We verify the effectiveness of proposed models with objective and subjective tests. AXY score over GST is 2.01 and 1.14 for hard pitchtron and soft pitchtron respectively.

## 1. Introduction

Audiobook has two requirements to meet. First, there are requirements for the reading voice. Listeners want it to be from their favorite celebrities, or even their family members who are not professional voice actors. Also, it needs to be rich in prosody so that it will not bore the listeners. There are hardships in meeting these requirements. To begin with, it is impractical to obtain voice of speakers who are not professionals for more than 30 minutes. Secondly, when asked to read scripts for collecting DB, ordinary people feel uncomfortable voice acting and it ends up sounding monotonous and boring, which is not a desirable quality to generate audiobooks. This is where the prosody transfer based on multi-speaker TTS comes in. Prosody transfer is the technique to transfer desired speaking style of reference audio into target speaker's voice. It is considered parallel prosody transfer if the content of reference audio and target sentence being synthesized must match, otherwise, non-parallel. Although global style token (GST) proposed by Wang et al. serves its role of unsupervised, non-parallel prosody transfer, there are room for improvements [1]. First, the process of exploring the role of each token is exhausting, not to mention, assigning the right value for style token weights, since the embedding space is nonlinear. Secondly, this approach is inappropriate for the case where training data are not so rich in prosody because they are collected from multiple non-professional speakers. Not just GST, numerous variational autoencoder (VAE) based prosody transfer approaches also require the training DB to contain rich speaking styles so that the probability distributions can be learned by the model [2-8]. However, we aim to generate speech in prosody as rich as that of professional voice actors while using corpus from ordinary people who are not professional voice actors. The former has wider vocal, energy and pausing length range, while the latter has narrower one. Transferred sound has glitches if prosody is transferred across people from different vocal ranges, energies and dialects. This mismatch is hard to fine-tune for the target speaker with GST since the token weight scaling is unpredictable. To address above issues, we suggest two prosody transfer models and call them *hard pitchtron* and *soft pitchtron*, respectively. It is named as pitchtron because we exploit the pitch information together with Tacotron2 decoder so as to resolve aforementioned issues. Main contributions of our work are as follows.

- We propose two prosody transfer models with higher transferability to prosodies that are rarely or never seen during training.
- We release a toolkit 'Pitchtron'[1] and Korean DB 'emotion-tts'[2].

## 2. Related works

There have been many works on prosody transfer and expressive speech synthesis. For the supervised approaches, researchers utilize labels that annotate the accent type or linguistic description of the text [9-12]. There is also semi-supervised approach. Koriyama et al. attempts to use Gaussian process latent variable model to represent pitch contours of a tonal language [13]. Unsupervised training for the prosody modeling is also widely being studied because labeling is costly and the annotation can be subjective. Skerry-Ryan et al. defined prosody as the residual attributes of the speech that remain after accounting for speaker and text [14]. They use autoencoder to encode such residual attributes of speech. Wang et al. suggested decomposing the residual embedding with style tokens and obtained somewhat disentangled control over prosody attributes in an inductive manner [1]. Lee et al. suggested using variable-length residual embedding to improve fine-grained control over local dynamics [15]. Studies on unsupervised prosody control continues in order to achieve disentangled control over prosody attributes. There are roughly three directions of researches to obtain this goal. The first is to use prosody parameters such as F0, MGC to the autoencoder instead of using Mel spectrogram which is 'entangled' acoustic feature. Klimkov et al. and Gururani et al. use prosody features to train autoencoder [3,16]. Valle et al. use F0 as input to decoder [17]. This inspired our study that is further described in section 3. The second is to use VAE. VAE can disentangle its latent variable dimensions since it assumes diagonal covariance matrix for prior distributions. The downside is, even though 'label' of certain prosody is not required, the utterances of desired prosody needs to be included in the training data to some amount that is enough for the model to learn the distribution. Akuzawa et al. and Zhang et al. use VAE module to sample prosody embedding and use this as condition to the VoiceLoop decoder and Tacotron2 decoder, respectively. [4, 5, 18, 19]. Wan et al. uses hierarchical approach to encode prosody using linguistic context and VAE [6]. Hsu et al. assumes hierarchy in acoustic features [7]. Thirdly, gradient reversal technique also helps to disentangle speech attributes. Zhang et al. adopts gradient reversal layer (GRL) between speaker classifier and text embedding in order to disentangle speaker characteristic from text embedding [20]. Hsu et al. also adopts GRL between augmentation classifier and speaker embedding as they attempt to disentangle noise property and speaker characteristic in order to train multi-speaker TTS with low-SNR data [21].

---

[1] https://github.com/hash2430/pitchtron
Sound demos can be found at
https://sunghee.kaist.ac.kr/entry/pitchtron

[2] https://github.com/emotiontts/emotiontts_open_db

## 3. Proposed methods

### 3.1. Hard pitchtron

Figure 1 shows hard pitchtron. We use F0s as decoder input as well as mel spectrograms as in [17]. Due to its dependency on reference F0 sequences, non-parallel prosody transfer is limited to sentences with similar sentence structure as the reference audio. GRL between prosody embedding and speaker classifier is adopted to solve the problem where certain speaker always speaks in certain style. For example, dialect speaker always speaks in dialect and we need to disentangle prosody from speaker in order to enable other speakers to speak in that dialect. In equation (1), prosody embedding $r$ is fed to GRL then fed to speaker classifier to predict speaker embedding $s$. The GRL does nothing for forward step but multiplies -1 to the gradient during backward step. Hard pitchtron is trained with equation (2) as objective function and we experimented with various values for $\lambda$. By unlearning the cross-entropy term in equation (2), we are forcing the prosody encoder to extract prosody embedding that is oblivious of the speaker. In equation (2), $N$, $Y$, $g$ and $s$ are batch size, predicted mel spectrogram, gate decision and speaker. $\hat{Y}$, $\hat{g}$ and $\hat{s}$ are ground truth counterparts. $RMSE$, $BCE$ and $CE$ are root mean square error, binary cross entropy and cross entropy, respectively.

$$s = Classifier(GRL(r)) \quad (1)$$

$$L = \frac{1}{N}\sum_{i=1}^{N}[RMSE(Y,\hat{Y}) + BCE(g,\hat{g}) + \lambda \cdot CE(s,\hat{s})] \quad (2)$$

### 3.2. Soft pitchtron

Figure 2 shows soft pitchtron. Instead of using mel-spectrogram as input to the prosody encoder, this model only uses pitch contour. This results in significant performance gains in MOS and prosody similarity, which will be discussed in section 4. On top of that, this model is capable of non-parallel prosody transfer where there are no restrictions about target sentences to be generated. That is because this model aligns target text with reference audio using attention mechanism as in the work of Lee et al. [15]. The difference is, in soft pitchtron, you do not need to analyze each dimension of prosody embeddings to find out which dimension should be manipulated to control pitch. Instead, you only need to control the input pitch sequence to fit the target speaker vocal range or to manipulate the pitch contour on temporal axis. Not only that, the scale and direction of manipulation is transparent and predictable because we are scaling the input in F0 space, not in embedding space.

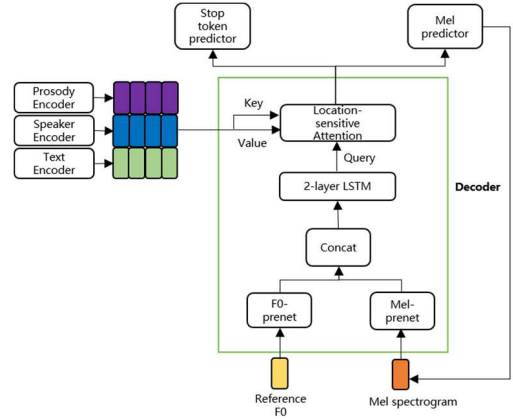

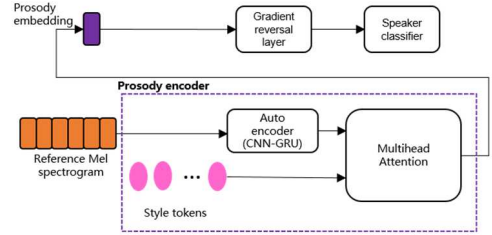

Figure 1: *Block diagram of hard pitchtron.*

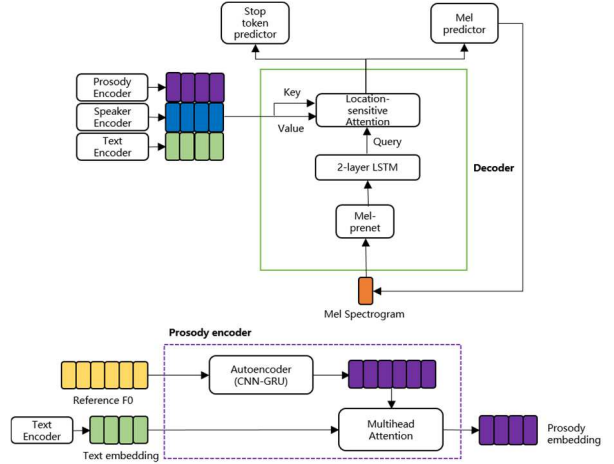

Figure 2: *Block diagram of soft pitchtron.*

## 4. Experiments

### 4.1. DB

#### 4.1.1. Training set

In our study, 1) 13 hours of speech from female professional single speaker who is not so expressive in style and 2) 93 hours of speech from 158 speakers are used. 129 people in this multi-speaker DB are non-professional and thus, it is rather monotonous and they mostly have limited vocal range. We find this setting realistic to have large size low-quality multi-speaker DB without much style variance from ordinary people. Also, having single speaker with large corpus was essential for the model to learn the alignment during initial training epochs. The single-speaker part is included in emotion-tts DB which we are releasing with this paper. The multi-speaker part consists of publicly available Korean DB[22] and emotion-tts. Both DB are gender-balanced and age-balanced, ranging from 20s to 70s and 10s to 50s respectively. Also, the latter includes two dialect speakers who speak 700 sentences of Kyoungsang and Cheolla dialects respectively. We use 678 utterances of Kyoungsang dialect and none of Cheolla dialect for training to test the generalizability to unseen prosody.

*4.1.2. Test set*

We set aside two utterances for each training speaker for test. Except, 20 utterances are set aside for Kyoungsang dialect speaker because GST made many attention mistakes and we had to pick utterances that had reasonable alignments. Since FFE and MCD can be affected by vocal range difference of target and reference speaker, we do the objective test with matching target speaker and reference speaker. For subjective test, emotive utterances and Cheolla dialect utterances which are never seen during training are additionally tested as reference audio along with prosodies in the training set. Emotive utterances are also part of emotion-tts DB. For subjective tests, reference speakers are all unseen during training, which we believe to be a realistic assumption.

### 4.2. Baseline

As a baseline to compare with proposed methods, we implemented and experimented with GST model. As expected, GST model deteriorated under the setting of our study, where training DB and target speakers are mainly ordinary people of narrow vocal range and narrow prosody variation, while references have bigger dynamic range. Cracking sounds have been frequently observed where the reference audios were out of target speaker's vocal range. Although it is possible to scale the vocal range by manipulating token weights, fine-tuning for each target speaker is impossible since the scale is non-linear and unpredictable in the embedding space. Also, since reference speakers who are professional had voices with much higher energy, the token weight representing energy peaked during synthesis, resulting in annoying sound. Plus, when given dialect that is unseen during training as reference, the token that represents pause length peaked and generated abnormally long pause. These unusual token weights also resulted the attention alignment to collapse. These abnormal token weight values are resulted from the fact that we tested with reference styles that are barely or never seen during training stage, which is different assumption from the original paper where GST is proposed [1].

### 4.3. Features and training setup

We follow the settings of GST [1], otherwise mentioned. Inputs are phonemes. WaveGlow vocoder is used for all models. 22,050 Hz sampling rate, 1,024 window length, 256 hop size, 10 GST tokens are used. Adam optimizer is used with initial learning rate of 1e-3 and learning rate decayed into half per 50,000 steps.

### 4.4. Objective evaluations

Table 1 and Table 2 show objective evaluation for baseline and proposed models on standard Korean dialogic style and Kyoungsang dialect, respectively.

Table 1: *Objective evaluations for standard Korean.*

|  | GPE (%) | VDE (%) | FFE (%) | MCD (dB) |
|---|---|---|---|---|
| GST | **9.33** | **2.53** | **8.17** | **5.85** |
| Ps | 6.85 | 1.66 | 6.36 | 4.80 |
| Ph ($\lambda = 0.0$) | 1.97 | 1.57 | 2.94 | 3.76 |
| Ph ($\lambda = 0.02$) | **0.54** | **1.50** | **1.88** | **3.48** |
| Ph ($\lambda = 0.2$) | 0.74 | 1.57 | 2.08 | 3.62 |
| Ph ($\lambda = 2.0$) | 2.82 | 1.64 | 3.57 | 4.08 |

Table 2: *Objective evaluations for Kyoungsang dialect.*

|  | GPE (%) | VDE (%) | FFE (%) | MCD (dB) |
|---|---|---|---|---|
| GST | **26.81** | **2.91** | **18.48** | **6.13** |
| Ps | 19.59 | 3.39 | 15.13 | 5.79 |
| Ph ($\lambda = 0.0$) | 4.50 | 3.20 | 6.05 | 3.56 |
| Ph ($\lambda = 0.02$) | 2.71 | 2.26 | 3.97 | 3.30 |
| Ph ($\lambda = 0.2$) | **2.69** | **1.98** | **3.69** | **3.34** |
| Ph ($\lambda = 2.0$) | 4.11 | 3.00 | 5.62 | 3.63 |

'GST' represents global style token, 'Ps' represents soft pitchtron and 'Ph' represents hard pitchtron. 'GPE' is gross pitch error and 'VDE' is voicing decision error. For both types of prosodies, soft pitchtron and hard pitchtron gave lower F0 frame error (FFE) and Mel cepstral distortion (MCD) than GST. Especially the error rate reduction of hard pitchtron compared to GST is remarkable for the Kyoungsang dialect. Kyoungsang dialect is pitch-accented and hard pitchtron benefits from its nature of exploiting pitch information. We also experimented with different weight values for gradient reversal loss term. For both prosodies, having GRL in between speaker classifier and prosody embedding significantly improves the FFE and MCD. For standard Korean prosody, the weight for GRL did not have to be so big and 0.02 gave the best result. On the other hand, for the Kyoungsang dialect, 0.2 gave the best result. We think that this loss term plays important role especially for the Kyoungsang dialect because there is only one speaker who speaks this dialect and thus disentangling the speaker embedding and prosody embedding is essential for transferring Kyoungsang dialect. However, if we further increased the weight value, the loss term for mel spectrogram construction is relatively compensated and FFE and MCD increased. Figure 3 demonstrates qualitative analysis of prosody transferability and vocal range controllability for proposed pitchtron models and GST. By comparing (a), (c), and (e), it can be seen that hard pitchtron is very powerful at transferring the reference audio while soft pitchtron and GST are comparable. Figure 3. (b), (d), and (f) show pitch controllability which is useful for both fitting reference audio into the vocal range of target speaker and manipulation of pitch on the time axis to fine-tune certain nuance. The latter use case is only possible for hard and soft pitchtron since they use variable length embeddings unlike GST. (b), (d) are plotted by simply scaling the reference pitch contour by half for all time steps. (e) is plotted by analyzing each tokens of GST in order to find the token in charge of pitch scale and experimenting with diverse weight values for that token to figure out the right scale in the embedding space. It can be seen that hard pitchtron has very strict control over pitch scaling while (d) and (e) are not so strictly following the desired scale. Still, (d) has its advantage of easy controllability over (e). Figure 4 shows transferability of soft pitchtron and GST for non-parallel sentences. Even though there are no restrictions for the difference between reference and target sentences, they are chosen to be of similar length for clearer visualization. It is observed that soft pitchtron and reference audio both have pitch contour that tops at the middle of sentence and goes down, while such pattern is unclear for the GST model.

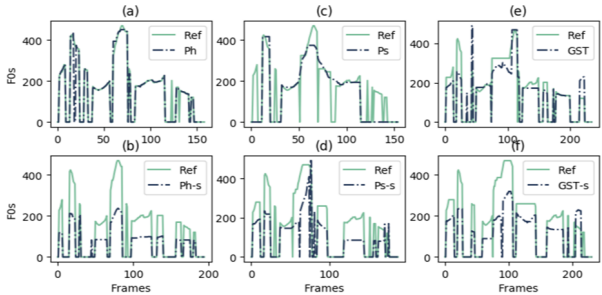

Figure 3: *(a) Pitch transferability of hard pitchtron. (b) Scalability of hard pitchtron. (c) Pitch transferability of soft pitchtron. (d) Scalability of hard pitchtron (e) Pitch transferability of GST (f) Scalability of GST.*

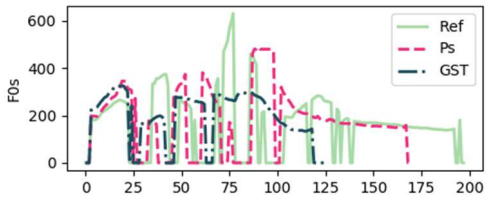

Figure 4: *Non-parallel prosody transfer.*

### 4.5. Subjective evaluations

All subjective tests on this section are done by using unseen reference speaker, thus reference speaker and target speaker are different. We believe this is a realistic assumption. This setting makes the vocal range scalability of proposed methods crucial. 34 native Korean speakers participated in this evaluation. Table 3 shows AXY prosody transfer test result. Subjects were asked to evaluate prosody resemblance with reference audio A in scale of -3 to 3. Negative sign is when subject preferred GST over pitchtron. 0 is when they had no preference among GST and the proposed. Positive sign is when pitchtron is preferred. The placement of GST and pitchtron as X and Y are random at the test time to prevent bias. First row is preference of hard pitchtron over GST and second row is of soft pitchtron over GST. As expected, hard pitchtron exhibits significant preference. What is interesting is that subjects were asked to decide their preference not just by pitch similarity, but by overall prosody factors but they still showed high preference for both pitchtron models over GST. Originally, we expected the preference will only be strong for pitch-accented dialects such as Kyoungsang and Cheolla, but the preference was observed over all 4 types of prosodies. 'Dialogic' and 'Emotive dialogic' are in standard Korean and it is not a pitch-accented language.

We think that this proves the importance of pitch contour on the perception of overall prosody. Table 4 shows MOS of ground truth (T), GST (G), hard pitchtron (H) and soft pitchtron (S) for dialogic, emotive dialogic, Kyoungsang and Cheolla prosodies. Ground truth MOS is rather low due to public DB. What is noticeable here is that GST consistently gives low MOS for all prosodies. That is the deterioration pointed out earlier in 4.2. due to the vocal range, energy and pause length mismatch between training data and reference audio given at inference stage. Mismatch due to vocal range is effectively solved for both hard pitchtron and soft pitchtron, thus resulting in better MOS than GST. Also, for soft pitchtron, MOS was even higher because as this model only uses pitch for the prosody embedding, the energy level was kept constant. Even though this restricts prosody similarity to some degree, this made synthesized speech more comfortable to listen to, affecting positively in terms of MOS.

Table 3: *Preference score for prosody transfer.*

|   | Dialogic | Emotive | Kyoungsang | Cheolla |
|---|---|---|---|---|
| H | 1.94±0.15 | 1.82±0.15 | 1.95±0.14 | **2.01±0.15** |
| S | 0.84±0.13 | **1.14±0.12** | 0.96±0.14 | 0.78±0.13 |

Table 4: *MOS evaluation for speech quality.*

|   | Dialogic | Emotive | Kyoungsang | Cheolla |
|---|---|---|---|---|
| T | 3.83±0.11 | | | |
| G | 2.35±0.11 | 2.28±0.12 | 2.12±0.11 | 2.51±0.12 |
| H | 3.32±0.10 | 3.34±0.12 | 3.02±0.12 | 3.30±0.11 |
| S | 3.34±0.10 | 3.73±0.12 | 3.36±0.11 | 3.36±0.11 |

## 5. Conclusions

In this paper, we proposed two prosody transfer models that are better suited for realistic conditions where training dataset is composed mainly of ordinary people's voice and each person have about 30 minutes of utterances while reference audios are given to be rich in prosody and made by professional voice actors. We discovered that controllability over pitch was the key to resolving the vocal range mismatch and removing cracked sound and other glitches coming from the mismatch of training data and reference audio. We proposed two models hard and soft pitchtron which use pitch as decoder input and prosody encoder input, respectively. Hard pitchtron gave 2.01 preference of prosody transferability over GST at most. While soft pitchtron gave 1.14 AXY score which is less impressive than hard pitchtron, but it had impressive improvement on MOS compared to GST. For future work, we are planning on conditioning TTS decoder on emotion embedding extracted directly from text instead of reference audio to generate expressive speech.

## 6. Acknowledgements

This material is based upon work supported by the Ministry of Trade, Industry & Energy (MOTIE, Korea) under Industrial Technology Innovation Program (No. 10080667, Development of conversational speech synthesis technology to express emotion and personality of robots through sound source diversification).